# Electrically Sign-Reversible Topological Hall Effect in a Top-Gated Topological Insulator (Bi,Sb)$_2$Te$_3$ on a Ferrimagnetic Insulator Europium Iron Garnet

Jyun-Fong Wong,[1,°] Ko-Hsuan Mandy Chen,[1,°] Jui-Min Chia,[1] Zih-Ping Huang,[2] Sheng-Xin Wang,[1] Pei-Tze Chen,[1] Lawrence Boyu Young,[2] Yen-Hsun Glen Lin,[2] Shang-Fan Lee,[3] Chung-Yu Mou,[1,4] Minghwei Hong,[2,*] and Jueinai Kwo[1,*]

[1]Department of Physics, National Tsing Hua University, Hsinchu 30013, Taiwan

[2]Graduate Institute of Applied Physics and Department of Physics, National Taiwan University, Taipei 10617, Taiwan

[3]Institute of Physics, Academia Sinica, Taipei 11529, Taiwan

[4]Center for Quantum Technology, National Tsing Hua University, Hsinchu 30013, Taiwan

[°]J.-F. W. and K.-H. M. C. contributed equally to this work.

[*]Address correspondence to J. Kwo, raynien@phys.nthu.edu.tw; M. Hong, mhong@phys.ntu.edu.tw

ABSTRACT

Topological Hall effect (THE), an electrical transport signature of systems with chiral spin textures like skyrmions, has been observed recently in topological insulator (TI)-based magnetic heterostructures. However, the intriguing interplay between the topological surface state and THE is yet to be fully understood. In this work, we report a large THE of ~10 Ω (~4 μΩ·cm) at 2 K with an electrically reversible sign in a top-gated 4 nm TI $(Bi_{0.3}Sb_{0.7})_2Te_3$ (BST) grown on a ferrimagnetic insulator (FI) europium iron garnet (EuIG). Temperature, external magnetic field angle, and top gate bias dependences of magnetotransport properties were investigated and consistent with a skyrmion-driven THE. Most importantly, a sign change in THE was discovered as the Fermi level was tuned from the upper to the lower parts of the gapped Dirac cone and *vice versa*. This discovery is anticipated to impact technological applications in ultralow power skyrmion-based spintronics.

Three-dimensional topological insulators (TIs), featured with their spin-momentum-locked topological surface states (TSSs),[1,2] have generated enormous interest in spintronics over the past decade. The interplay between the spin-momentum-locked TSS and magnetism results in novel electrical transport phenomena.[3] A well-known example is the observation of quantum anomalous Hall effect (QAHE) in a transition metal-doped TI (Bi,Sb)$_2$Te$_3$ (BST).[4-6] Besides the magnetically doped TIs, much attention is also given to TI heterostructures interfaced with magnetic materials (MMs) to attain a long-range magnetic ordering *via* magnetic proximity effect (MPE).[7-14] However, most studies were focused on the discussion of the phenomena arising from the non-trivial Berry curvature in reciprocal space. The investigation of the spin textures in real space and their related electrical transport in TI heterostructures remains largely unexplored.

Topological Hall effect (THE), a Hall response that emerges from the deflection of charge carriers flowing through non-trivial chiral spin textures, is a transport signature commonly used to identify these chiral spin textures, such as magnetic skyrmions.[15,16] Apart from the magnetic crystals lacking spatial inversion symmetry (SIS), such as B20-type chiral magnets,[17-20] skyrmions have been observed in SIS-broken magnetic heterostructures consisting of materials exhibiting strong spin-orbit coupling (SOC), such as heavy metals (HMs) or TIs.[21,22] The interfacial Dzyaloshinskii–Moriya interaction (DMI) plays a crucial role in stabilizing skyrmions in these SIS-broken magnetic heterostructures.[23,24] In contrast to the high current density required for magnetic domain wall motions ($10^{11}$–$10^{12}$ A/m$^2$), the ultralow current density ($10^5$–$10^6$ A/m$^2$) achieved in skyrmion motions brings great potential in energy-efficient devices for memory and computing technology.[16,20,23,25] THE, being an electrical transport phenomenon, is a promising method of skyrmion detection in the field of skyrmion-electronics.[24,26]

Among many THE studies, there have been debates about the exact origin of the reported Hall responses.[27-33] Because THE is commonly intertwined with anomalous Hall effect (AHE), an alternative cause is the superposition of multi-AHE contributions.[27-29,31-33] In this scenario, one can conduct two- or three-AHE fits to decompose the signals, and ascertain if the THE responses could indeed result from the overlapping of multi-AHE components.[14,30,34,35] Although several reports demonstrated pronounced THE in TI-based heterostructures,[30,34,36-41] it is noteworthy that an in-depth discussion on the THE emerging from the carrier transport of spin-momentum-locked TSS is still lacking. A straightforward way to investigate the relationship between THE and charge carriers in TSS is by implementing an electrical gate bias on TI. So far, the gate-tunable THE has been reported in Mn-doped Bi$_2$Te$_3$ films and Cr-doped BST heterostructures;[37,40,41] however, the gate devices were achieved by using SrTiO$_3$ (STO) as a dielectric with a large gate bias of tens of volts,[37,40,41] which are unfavorable in practical use. Finding a suitable gate dielectric with excellent carrier tunability and reliability remains an important issue and needs to be addressed for both fundamental scientific studies and technological applications.

In this work, we report the observation of THE in TI BST/ferrimagnetic insulator (FI) europium iron garnet (EuIG) bilayers. In particular, we successfully manipulated the amplitude and the sign of THE by altering the carrier density and carrier types using a top electrical gate within a few volts, which has yet to be reported. By adopting a heterostructure of gate oxide/TI/FI, the current flow path and the resulting Hall contribution can be limited to the TI layer, which is rather simple compared to other magnetic TI (MTI)/TI/MTI or TI/MM heterostructures. Judging from the temperature, magnetic field angle, and gate voltage dependences of Hall measurements, our findings were consistent with a skyrmion-driven THE instead of a superposition of AHE loops. The largest THE magnitude reached ~10 Ω (~4 μΩ·cm) under an applied gate voltage ($V_g$) of -0.6 V at 2 K. Most importantly, the pronounced and repeated THE sign reversal occurs when the Fermi level ($E_F$) is tuned across the gapped Dirac point (DP) of BST, suggesting a unique THE characteristic related to the TSS. Hence, the exploitation of THE in TI-based heterostructures opens a new route in high-density and ultralow-power skyrmion-based devices in spintronics.

The material growth of BST/EuIG followed the previous work of our group,[14,42] and physical characterizations on the sample are presented in Section S1 in Supporting Information. Briefly, FI EuIG(001) films with perpendicular magnetic anisotropy were grown on GGG(001) by off-axis magnetron sputtering. BST was then deposited on EuIG using molecular beam epitaxy with Bi:Sb

= 3:7 to reach a bulk insulating TI. The gate oxides were made of e-beam evaporated $Y_2O_3$ and atomic layer deposition-grown $Al_2O_3$. The samples were patterned into Hall bars using photolithography. Electrical transport measurements were conducted on a 4 nm BST thin film grown on a 20 nm EuIG with simultaneously magnetized top and bottom TSSs in a Quantum Design physical property measurement system.

Figure 1a displays the Hall resistance of the ungated 4 nm BST/EuIG at 2 K after the subtraction of the linear background from the ordinary Hall effect (OHE). In addition to the MPE-induced AHE loop following the square-shaped magnetic hysteresis of EuIG,[14,42] excessive Hall signals were clearly observed over a wide range of magnetic fields from -1.5 to 3 kOe as the magnetic field was swept from negative to positive. These excessive Hall signals reached the largest value near the coercive field ($H_c$) and behaved like THE. This THE-like feature has been repeatedly observed in many BST/EuIG samples previously.[14] Notice that it is possible to have artificial THE-like responses from two overlapping AHE contributions.[27-33] Here, we performed two additional experimental checks, the minor loop approach and AHE curve fittings, to resolve this issue. In the former method, the Hall traces coincided under successively increasing sweeping magnetic fields as the expected behavior of THE responses without other AHE loops.[32] (see Section S2 in Supporting Information) In the latter method, although the Hall signals could be well fitted with two AHE contributions mathematically, it is rather difficult to reconcile their dependences on temperatures as well as gate biases with a reasonable or existing physical picture. (see Section S3 in Supporting Information) Therefore, we infer that these excessive Hall signals in BST/EuIG result from a genuine THE.

After excluding the Hall signals from the superposition of AHE loops, we proceeded to extract the THE resistance ($R_{THE}$) by subtracting the AHE resistance ($R_{AHE}$). In view of the square magnetic hysteresis observed in the alternating gradient magnetometer measurement of BST/EuIG at 300 K, and square MPE-induced AHE loops at temperatures without the existence of THE, we here fit the AHE component using $R_{AHE} = R_{AHE-max} \tanh\left(\frac{H \pm H_c}{H_0}\right)$, where $R_{AHE-max}$ and $H_0$ are the maximum of the $R_{AHE}$ and fitting parameter, respectively.[30,34] Note that how we determine the AHE contribution could affect the extracted THE components. The rationale for subtracting a single square AHE loop is given in Section S4 in Supporting Information. After the subtraction of the square AHE loop (black solid line in Figure 1a), the $R_{THE}$ data is shown in Figure 1b; the maximum of the $R_{THE}$ is denoted by $R_{THE}^{MAX}$, and the $H$ field corresponding to the $R_{THE}^{MAX}$ is denoted by $H_T$. A giant $R_{THE}^{MAX}$ of ~8 Ω was observed without an applied $V_g$ at 2 K. At the applied $V_g$ of -0.6 V, the $R_{THE}^{MAX}$ reached the largest value of ~10 Ω (~4 μΩ·cm). The THE magnitude in this work is the second largest among the reported TI-based bi-layer systems.[30,34,36,38,39] (See Table S1 in Supporting Information)

Next, we discuss the temperature dependence of THE in BST/EuIG. As presented in Figures 1c, d and S5 in Supporting Information, the THE gradually diminished with increasing temperatures and disappeared at 75 K, which could be attributed to the thermal fluctuation or the reduced DMI strength, consistent with the previous studies in TI-based systems.[14,30,34,36-38,40] Furthermore, shown in the color map of Figure 1c is the temperature dependence of $H_T$ that follows closely with that of $H_c$. This observation is in accord with reports on the THE driven by the magnetization reversal process in systems hosting magnetic skyrmions.[34,43] It is noteworthy that our THE were also found at zero fields, suggesting a robust skyrmion phase without the support of an external magnetic field, similar to the observation in FeGe.[44,45] The $R_{THE}$ at zero fields ($R_{THE}^0$) showed an akin temperature dependence to that of $R_{THE}^{MAX}$ and also vanished at 75 K, as demonstrated in Figure 1d. The critical temperature for observing the THE in this work is notably higher than those reported in Cr-doped BST/BST (18 K),[36] $Cr_2Te_3/Bi_2Te_3$ (20 K),[39] and BST/MnTe (20 K),[38] but slightly lower than $Bi_2Se_3/BaFe_{12}O_{19}$ (80 K) and $CrTe_2/Bi_2Te_3$ (100 K).[30,34] (See Table S1 in Supporting Information)

To deepen our understanding of the THE in BST/EuIG, we further investigated the angular dependence of the Hall effect data with the measurement geometry illustrated in Figure 2a. The angle $\theta$ is defined as the angle between the external magnetic field $H$ and the surface normal direction +z in the x-z plane. The Hall traces at 5 K after subtracting the linear OHE background from 10° to 80° are demonstrated in Figure 2b. The AHE loops expanded with increasing $\theta$. The $H_c$ enlarged and was proportional to 1/cos$\theta$ (white dashed line in Figure 2c), indicating that a larger external magnetic field was needed to flip the magnetization in BST to the opposite direction as the $\theta$ increased. Furthermore, significant THE responses coexisting with AHE were observed and sustained to 70°. The $R_{THE}$ as a function of $\theta$ and $H$ field is summarized in the color map of Figure 2c, as the $H$ field was swept from negative to positive. Similar to the temperature dependence of $H_T$ and $H_c$ discussed in the previous paragraph, the angular dependence of $H_T$ also followed closely with that of $H_c$ before the disappearance of THE.

To better elucidate the angle-dependent results, the magnitude of THE was discussed as a function of the *in-plane* component of the applied magnetic field. Because the $H_T$ closely followed the trend of $H_c$ before the disappearance of THE, the $H_T$ was approximately proportional to 1/cos$\theta$, indicating that the *out-of-plane* component of the applied field ($H$ cos$\theta$) was nearly the same when the $R_{THE}^{MAX}$ occurred. Therefore, we plotted the $R_{THE}^{MAX}$ as a function of the *in-plane* magnetic field $H_T$ sin$\theta$, as shown in Figure 2d. To give a reasonable estimate of the critical field for the disappearance of THE, $H_c$ sin80° was used for the data point of $\theta$ = 80°. Judging from Figure 2d, the $R_{THE}^{MAX}$ remained nearly the same with a moderate *in-plane* magnetic field (0.1–2.4 kOe), as attributed to the robustness of the topologically protected skyrmions and similar to the work of Peng *et al.*[30] The critical *in-plane* field for the disappearance of THE was between 2.4–5.3 kOe. The vanished THE above 5.3 kOe suggests the collapse of the skyrmionic state.[46-49] Similar angular behavior of THE was reported in the $Mn_{1-x}Fe_xSi$, a well-known B20-type chiral magnet hosting magnetic skyrmions, with the disappearance of THE at 55°.[46] (See Table S1 in Supporting Information for the comparisons with other materials) Hence, our observation of angle-dependent results in THE also supports the existence of skyrmions at the BST/EuIG interface.

In the following paragraph, we demonstrate the manipulation of THE by implementing a top electrical gate bias on BST. Figure 3a shows the gate dependences of longitudinal resistance ($R_{xx}$) and ordinary Hall coefficient ($R_H$). The $E_F$ was successfully tuned across the gapped DP of BST *via* the gate bias, as manifested by reaching a maximum in $R_{xx}$ and the sign change of $R_H$ near the charge neutrality point (CNP) (the gray-shaded region in Figure 3a). The CNP here is expected to locate at the center of the MPE-induced gap of TSS in BST. Moreover, the small applied $V_g$ for the CNP ($V_{CNP}$) of ~0.6 V indicated that the $E_F$ of our un-gated BST was close to the center of the magnetic gap, providing an excellent starting point to alter the carrier type and the carrier density of BST, thus to explore the THE behavior under an external electric field.

Figure 3b displays the gate dependence of selected AHE loops together with THE; the scattered points are the measured $R_{AHE}$ + $R_{THE}$ data, and the solid lines are the fitted AHE contributions. In the *p*-type region ($V_g < V_{CNP}$), a *hump* was found near the $H_c$ as the $H$ field was swept from negative to positive. (dark red arrows in Figure 3b) In the *n*-type region ($V_g > V_{CNP}$), a *dip* was observed instead near the $H_c$. (a green arrow in Figure 3b) With the applied $V_g \sim V_{CNP}$, the measured data became a square hysteresis loop, a typical feature of AHE without excessive Hall signals from THE. To extract the magnitude of THE, we subtracted the AHE contributions (black solid lines) and plotted $R_{THE}$ in Figure 3c. Positive THE was identified in the *p*-type region of the up-sweep (red) curves, while negative THE was observed in the *n*-type region. More data of gate-dependent THE loops with finer voltage steps are presented in Figure S6 in Supporting Information. The gate bias dependence of $R_{THE}^{MAX}$ is further summarized in Figure 3d.

To gain a better insight into the mechanism responsible for the sign reversal in THE, we examined $R_{THE}$ and its relations to other physical quantities. When a charge carrier flows through magnetic skyrmions, it will experience an effective electromagnetic field $B_{eff} = n_{sk} \Phi_0$, where $n_{sk}$ is the skyrmion density and $\Phi_0$ is the magnetic flux quantum.[15] The resulting Hall resistance is commonly approximated as:

$$R_{THE} \approx R_H P n_{sk} \Phi_0, \quad (1)$$

where $P$ is the local spin polarization of the conduction carrier following the spin texture of the skyrmion.[15,34] When the $E_F$ is tuned from the upper Dirac cone to the lower Dirac cone, the majority charge carriers will be altered from electrons to holes, leading to a sign change in $R_H$. In the meantime, the electron spin polarization is flipped to the opposite direction because of the spin-momentum locking of the gapped TSS. (see the Fermi surface in Figure 4) Since the propagation direction of holes is opposite with respect to that of electrons, the spin orientation for electrons and holes will thus be the same,[50] giving rise to the unchanged sign of the exchange coupling between charge carrier spins and local spin texture of skyrmion, *i.e.*, the same sign of $P$. Therefore, the sign of THE will be switched from negative to positive when the $E_F$ is varied from the upper to the lower parts of the Dirac cone, as illustrated in Figure 4. The $E_F$ modulation by the gate bias in Figure 3 was estimated roughly to be from ~84 meV ($V_g$ = +1.5 V) above the gapped DP to ~128 meV ($V_g$ = -1.5 V) below the gapped DP, as detailed in Section S8 in Supporting Information.

In the vicinity of CNP, namely $R_H \rightarrow 0$, the vanishing $R_{THE}$ could be attributed to the nearly equal numbers of *n*- and *p*-type charge carriers. The absence of the THE signals with the $E_F$ near the CNP was also observed in the Cr-doped BST sandwich heterostructures.[40,41] However, in their reports, $R_{THE}$ did not show a sign reversal with the applied gate voltages; the majority carrier type remained the same in both $V_g < V_{CNP}$ and $V_g > V_{CNP}$ regions, as suggested from the slopes of their Hall traces. In contrast, this work demonstrated an effective manipulation of charge carriers and THE *via* a top electrical gate. Moreover, our gate bias-dependent results lent strong support to a skyrmion-driven THE in the BST/EuIG heterostructure, which can be well described by Equation 1 for both positive and negative biases. The same behavior of THE manipulation *via* the top-gate bias was confirmed in another sample. (see Figure S7 in Supporting Information) In addition, the difference between the current work and the previous work of W.-J. Zou *et al.* was clarified and highlighted in Section S10 in Supporting Information;[14] the results are consistent among these two samples. Given these findings, inspecting the sign change in THE by tuning the $E_F$ across the CNP could be another viable way to differentiate the genuine and artificial THE in TI-based heterostructures. The electrically sign-reversible THE in TI/FI heterostructures could be a unique feature directly associated with the gapped TSS. This phenomenon has not been observed and might well be absent in other non-TI skyrmion systems, such as chiral magnets or HM/FI heterostructures like Pt/TmIG.

Finally, to further examine the reliability of the field-effect device and the reproducibility of the sign reversal in THE, we performed the THE measurements with a series of selected gate biases. As displayed in Figure 5a, the THE was reversibly switched. The $R_{THE}^{MAX}$ and $H_T$ values remained nearly identical to the starting ones, as demonstrated in Figures 5b, c. Although the repeated switching of this "THE" device is currently limited by the robustness of our gate oxides after applying excessive gate bias, the achievement of THE switching within only a few volts demonstrates tremendous progress in manipulating this phenomenon. The idea of an electrically tunable THE field-effect device may open up a new avenue in skyrmion-based spintronic applications.

In summary, we have demonstrated the electrically sign-reversible THE in a top-gated TI BST on a FI EuIG. The magnetotransport behaviors on temperature, external magnetic field angle, and gate bias were consistent with a skyrmion-driven THE, as opposed to the alternative mechanism of superimposed AHE loops. Moreover, the sign change in THE *via* an electrical gate could be a distinctive feature related to the gapped TSS in TI/FI compared to other non-TI skyrmion systems, such as chiral magnets and HM/FI heterostructures. Our findings are thus viable for an in-depth understanding of THE in a TI-based heterostructure. The BST/EuIG bilayer could serve as an excellent platform for studying the interplay among magnetism, chiral spin textures, and topological band structures. Especially, the reproducible electrical manipulation of THE within a few volts may be promising for ultralow-power skyrmion-related applications. Further investigations on direct imaging of skyrmion spin structures in real space will be essential, and experiments *via* scanning microscopy methods, such as spin-polarized scanning tunneling microscopy and scanning transmission x-ray microscopy, are now underway.

FIGURES

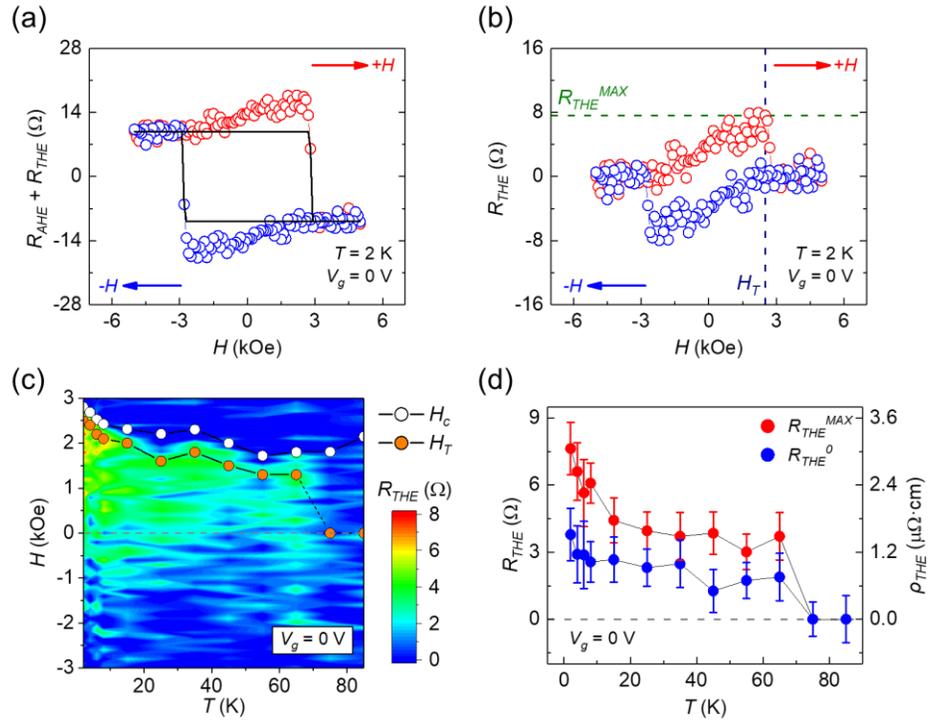

**Figure 1.** Temperature-dependent THE properties of BST/EuIG. (a) Coexistence of AHE and THE with an applied $V_g$ of 0 V at 2 K. The scattered points are the Hall data after subtracting a linear OHE background, and the solid line is the fitted AHE contribution. (b) The THE at $V_g = 0$ V. (c) Temperature dependences of $H_c$ and $H_T$, and a color map of $R_{THE}$ (color) as a function of temperature and magnetic field with an applied $V_g$ of 0 V; the $H$ field was swept from negative to positive. (d) The temperature dependences of $R_{THE}^{MAX}$ and $R_{THE}^{0}$.

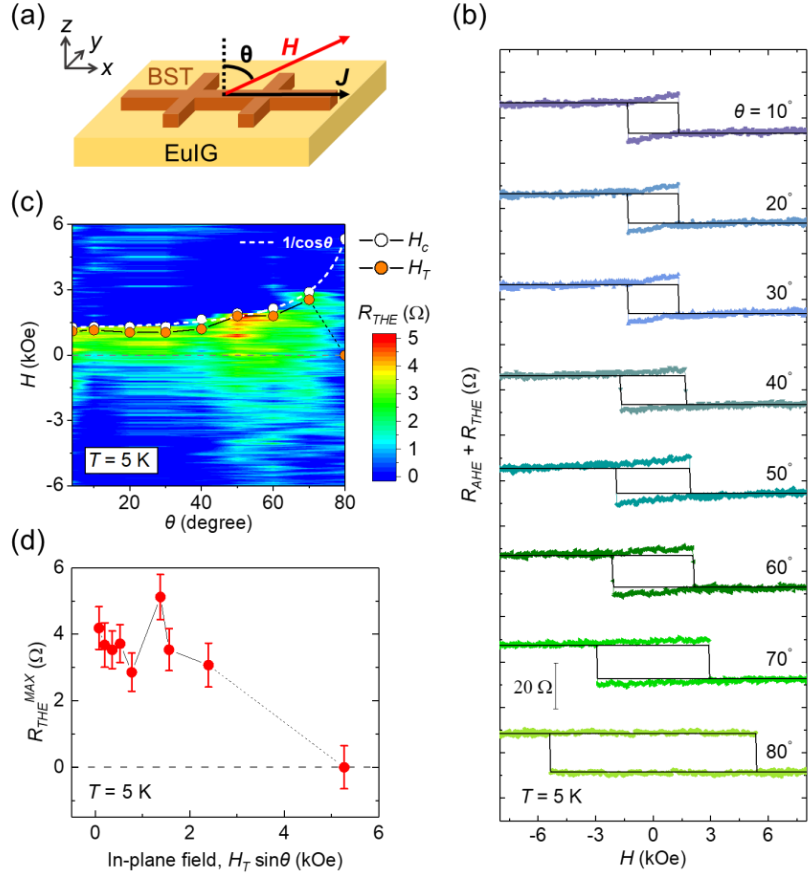

**Figure 2.** Angular dependence of THE from 10° to 80° at 5 K of BST/EuIG. (a) A schematic of the angle-dependent Hall measurements. (b) Angular dependence of the AHE loops plus the THE. (c) Magnetic field angle dependences of $H_c$ and $H_T$, and a color map of $R_{THE}$ (color) as a function of magnetic field angle and $H$ field; the $H$ field was swept from negative to positive. The white dashed line in (c) is the fitted curve of $H_c$ as a function of $1/\cos\theta$. (d) $R_{THE}^{MAX}$ as a function of *in-plane* magnetic field ($H_T \sin\theta$).

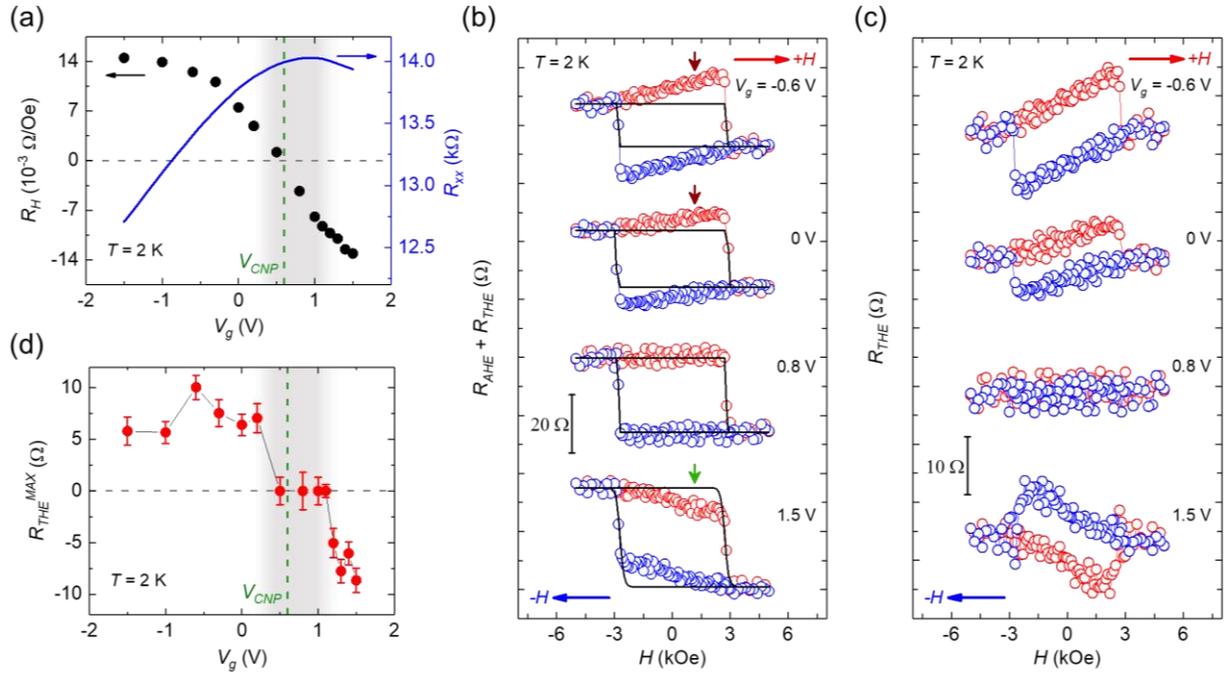

**Figure 3.** Gate-bias dependence of THE from -1.5 V to 1.5 V at 2 K of BST/EuIG. (a) Top-gate voltage dependences of $R_H$ (left) and $R_{xx}$ (right). Small $R_H$ and large $R_{xx}$ occurred in the vicinity of CNP (the gray-shaded region). (b) Concurrence of AHE and THE with selected gate voltages. The dark red and green arrows in (b) indicate the hump and dip features near the $H_c$, respectively. (c) THE with selected gate voltages. (d) Top-gate voltage dependence of $R_{THE}^{MAX}$ at 2 K. THE disappeared in the vicinity of CNP (the gray-shaded region).

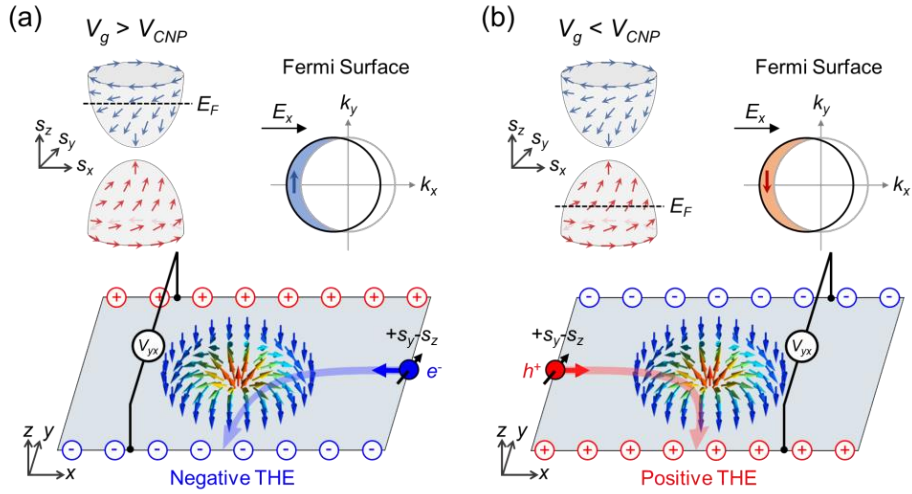

**Figure 4.** Illustration of the sign reversal in THE with varying gate bias. Schematics for gapped TSS, the Fermi surface with the application of current in +x direction, and the corresponding THE with (a) $V_g > V_{CNP}$ (*n*-type region, $E_F$ located at upper gapped TSS) and (b) $V_g < V_{CNP}$ (*p*-type region, $E_F$ located at lower gapped TSS), respectively.

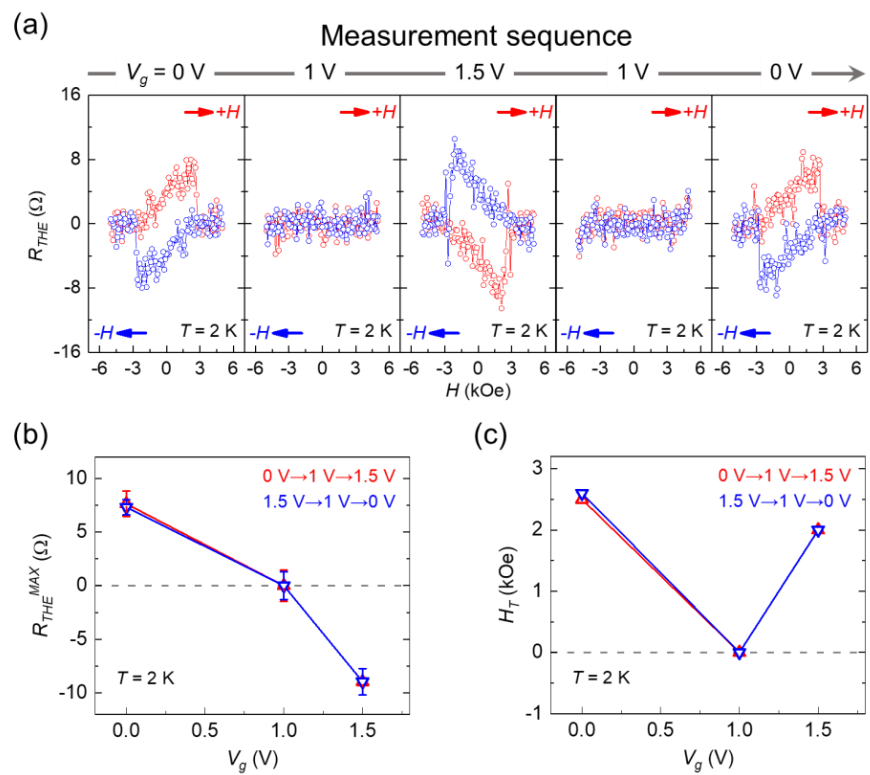

**Figure 5.** Repeated switching of THE in BST/EuIG. (a) THE loops with applied $V_g$ in the sequence of 0 V, 1 V, 1.5 V, 1 V, and 0 V. The corresponding modulation of (b) $R_{THE}^{MAX}$ and (c) $H_T$ as a function of $V_g$.

**ASSOCIATED CONTENT**

**Supporting Information**.

The Supporting Information is available free of charge at https://pubs.acs.org/XXX

Material growth and physical characterization of BST/EuIG heterostructures, Hall measurements using the minor loop method, discussion on the possibility of Hall effect data from two distinct AHE contributions and the rationale of subtracting a square hysteresis for the AHE loop to derive THE, table for the comparison of THE in various systems hosting magnetic skyrmions, temperature- and gate-dependent THE loops, estimate of $E_F$ manipulated by gate bias, gate-dependent results of one additional field-effect device, and clarification to accentuate the differences between this work and the previous gate-dependent THE results (PDF)

**AUTHOR INFORMATION**

**Corresponding Authors**

* **Jueinai Kwo** – *Department of Physics, National Tsing Hua University, Hsinchu 30013, Taiwan;* orcid.org/0000-0002-5088-6677; E-mail: raynien@phys.nthu.edu.tw

* **Minghwei Hong** – *Graduate Institute of Applied Physics and Department of Physics, National Taiwan University, Taipei 10617, Taiwan;* orcid.org/0000-0003-4657-0933; E-mail: mhong@phys.ntu.edu.tw

**Authors**

**Jyun-Fong Wong** – *Department of Physics, National Tsing Hua University, Hsinchu 30013, Taiwan;* orcid.org/0000-0001-5756-5801

**Ko-Hsuan Mandy Chen** – *Department of Physics, National Tsing Hua University, Hsinchu 30013, Taiwan;* orcid.org/0000-0002-4402-3541

**Jui-Min Chia** – *Department of Physics, National Tsing Hua University, Hsinchu 30013, Taiwan;* orcid.org/0000-0003-1759-4725

**Zih-Ping Huang** – *Graduate Institute of Applied Physics and Department of Physics, National Taiwan University, Taipei 10617, Taiwan;* orcid.org/0000-0001-8585-8812

**Sheng-Xin Wang** – *Department of Physics, National Tsing Hua University, Hsinchu 30013, Taiwan;* orcid.org/0000-0001-9725-2930

**Pei-Tze Chen** – *Department of Physics, National Tsing Hua University, Hsinchu 30013, Taiwan;* orcid.org/0000-0002-6249-5446

**Lawrence Boyu Young** – *Graduate Institute of Applied Physics and Department of Physics, National Taiwan University, Taipei 10617, Taiwan;* orcid.org/0000-0003-2569-6094

**Yen-Hsun Glen Lin** – *Graduate Institute of Applied Physics and Department of Physics, National Taiwan University, Taipei 10617, Taiwan;* orcid.org/0000-0002-0757-4109

**Shang-Fan Lee** – *Institute of Physics, Academia Sinica, Taipei 11529, Taiwan;* orcid.org/0000-0001-5899-7200

**Chung-Yu Mou** – *Center for Quantum Technology and Department of Physics, National Tsing Hua University, Hsinchu 30013, Taiwan;* orcid.org/0000-0003-2694-0992

**Author Contributions**

°J.-F.W. and K.-H.M.C. contributed equally to this work. J.-F.W. and J.-M.C. fabricated the Hall bar device, and collected and analyzed the transport data. Z.-P.H., S.-X.W., and P.-Z.C. produced the BST/EuIG samples. S.-X.W. conducted the magnetic measurements. L.B.Y. and Y.-H.G.L. deposited the *in-situ* top gate oxides. S.-F.L. and C.-Y.M. provided scientific inputs. J.K. and M.H. supervised the project. J.-F.W., K.-H.M.C., and J.K. composed and wrote the manuscript with the comments of all the authors.

**Notes**

The authors declare no competing financial interest.

**ACKNOWLEDGMENTS**

The authors would like to thank Professor Hsiu-Hau Lin, Professor Tay-Rong Chang, Keng-Yung Lin, Wei-Nien Chen, Wei-Jhih Zou, and Hsuan-Ning Chen for helpful discussions. This work was financially supported by the National Science and Technology Council (NSTC), Taiwan (NSTC 111-2112-M-007-043- and NSTC 111-2622-8-002-001-), and the Center for Quantum Technology and Department of Physics, National Tsing Hua University, Hsinchu 30013, Taiwan (NSTC 111-2634-F-007-006-). The authors acknowledge resources and support from the Quantum Materials Shared Facilities of Institute of Physics, Academia Sinica. The authors also thank the technical support from Taiwan Semiconductor Research Institute (TSRI), Taiwan.

For Table of Contents Only

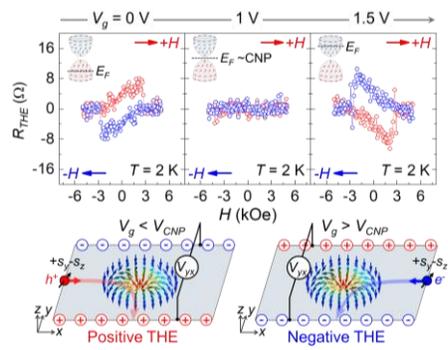

Supporting Information

# Electrically Sign-Reversible Topological Hall Effect in a Top-Gated Topological Insulator (Bi,Sb)$_2$Te$_3$ on a Ferrimagnetic Insulator Europium Iron Garnet

*Jyun-Fong Wong,[1]° Ko-Hsuan Mandy Chen,[1]° Jui-Min Chia,[1] Zih-Ping Huang,[2] Sheng-Xin Wang,[1] Pei-Tze Chen,[1] Lawrence Boyu Young,[2] Yen-Hsun Glen Lin,[2] Shang-Fan Lee,[3] Chung-Yu Mou,[1,4] Minghwei Hong,[2]\* and Jueinai Kwo[1]\**

[1]Department of Physics, National Tsing Hua University, Hsinchu 30013, Taiwan

[2]Graduate Institute of Applied Physics and Department of Physics, National Taiwan University, Taipei 10617, Taiwan

[3]Institute of Physics, Academia Sinica, Taipei 11529, Taiwan

[4]Center for Quantum Technology, National Tsing Hua University, Hsinchu 30013, Taiwan

°J.-F. W. and K.-H. M. C. contributed equally to this work.

\*Address correspondence to J. Kwo, raynien@phys.nthu.edu.tw; M. Hong, mhong@phys.ntu.edu.tw

**Contents**

**Section S1. Material growth and physical characterizations of BST/EuIG heterostructures**

20 nm thick FI EuIG(001) films with perpendicular magnetic anisotropy (PMA) were grown on GGG(001) at ~460 °C using an off-axis magnetron sputtering technique.[1] 4 nm thick TI BST films were then deposited on EuIG thin films at ~260 °C by using molecular beam epitaxy (MBE) under a base pressure below $4 \times 10^{-10}$ Torr.[2] The streaky reflection high energy electron diffraction (RHEED) pattern in Figure S1a manifested an atomically ordered and morphologically smooth BST surface with excellent crystallinity grown on EuIG thin films.

The deposition of gate oxides is described as follows. The samples were first transferred from the TI-MBE chamber to a multiple-chamber system containing oxide-MBE and atomic layer deposition (ALD) chambers under ultra-high vacuum (UHV) to avoid contaminations at the TI/oxide interface;[3] a 2 nm thick e-beam evaporated $Y_2O_3$ (in the oxide MBE chamber) followed by a 15 nm thick *in-situ* ALD $Al_2O_3$ was deposited on the pristine BST surface, leading to an unpinned $E_F$ at the interface of gate oxide/BST, based on our extensive expertise of high-κ dielectrics on semiconductor surfaces.[4-8] *Ex-situ* atomic force microscopy (AFM) was performed after the deposition of these oxides. The flat BST surface covered by the oxide layers with a low root-mean-square roughness of 0.633 nm is presented in Figure S1b. Next, the samples were patterned into Hall bars (880 μm × 90 μm) using photolithography. To better protect samples for transport measurements, a second $Al_2O_3$ layer was deposited with a thickness of 25 nm in another *ex-situ* ALD system after the fabrication of Hall bars to prevent current leakage from the edges of the Hall bar.

All electrical transport measurements were conducted in the physical property measurement system (PPMS) connected with the following instruments. A Keithley 2400 was used to provide a stable voltage source as the gate voltage. A Keithley 6221 was used as a current source to generate the alternating current with a root-mean-square amplitude of 100 nA. Two SR830 lock-in amplifiers were used as voltmeters to measure the Hall and longitudinal voltages, where the reference signal was provided by Keithley 6221. The Hall effect data of this work were antisymmetrized as a function of magnetic fields to eliminate the $R_{xx}$ component due to electrode misalignments. Figure S1c illustrates the device structure and the measurement setup.

Besides the Hall effect data discussed in the main text, we measured the $R_{xx}$ of BST/EuIG as a function of temperature for routine electrical characterizations, as shown in Figure S1d. $R_{xx}$ increased with decreasing temperatures from 300 K to 100 K, revealing a semiconducting behavior caused by the reduction of charge carriers in BST bulk.[9,10] Then, $R_{xx}$ reached a local maximum at ~100 K and decreased as the temperature was further lowered from ~100 K to ~10 K, indicating a metallic behavior from the TSS of BST.[9,10] At temperatures below 10 K, $R_{xx}$ increased with decreasing temperatures again, which could be attributed to the electron-electron interaction (EEI).[11] The longitudinal conductance ($G_{xx}$) was proportional to the natural logarithm of $T$, as shown in the inset of Figure S1d, consistent with the 2D EEI of the TSS.[12,13] As reported in the literature, the existence of the TSS could help generate a strong interfacial DMI, giving rise to topological magnetic structures with a pronounced THE.[14,15] In addition, the magnetic measurements were conducted in 4 nm and 7 nm thick BST/EuIG samples by alternating gradient magnetometer (AGM) at room temperature. A typical *out-of-plane* magnetic hysteresis is shown in Figure S2.

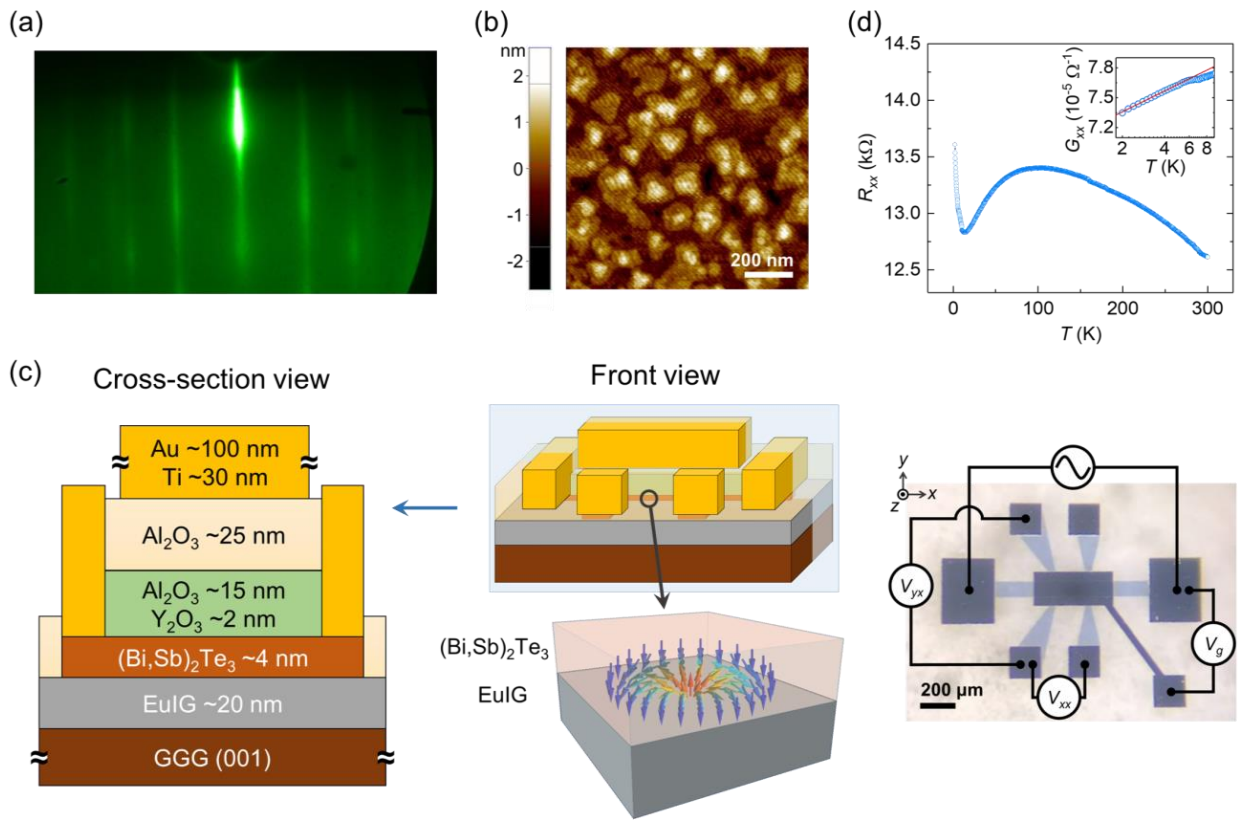

**Figure S1.** Sample characterization and schematic illustration of a top-gated BST/EuIG device. (a) RHEED pattern of the 4 nm BST(001) surface along the [100] axis. (b) Surface morphology of 15 nm ALD-$Al_2O_3$/2 nm e-beam evaporated-$Y_2O_3$/4 nm MBE-BST/20 nm sputtering-

EuIG/GGG(001) in a 1×1 μm² area by using AFM. (c) Schematics of a cross-section view and a front view of our top-gated BST grown on EuIG, in which a light blue vertical plane is to dissect the sample structure for the cross-section view. Also shown are the Néel-type skyrmion spin texture at the BST/EuIG interface, and the optical microscope image of a Hall bar device with our measurement setup. (d) $R_{xx}$ as a function of temperature. The inset in (d) shows the $G_{xx}$ as a function of temperature and the fit with a logarithmic function of $T$.

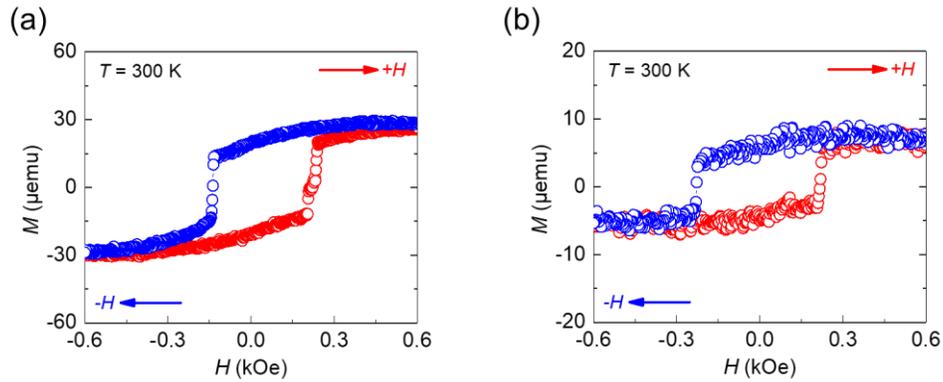

**Figure S2.** The *out-of-plane* magnetic hysteresis loop of (a) ~4 nm and (b) ~7 nm BST/EuIG.

**Section S2. Hall measurements using the minor loop method**

There has been debate if the chiral spin textures, such as skyrmion, can be identified *via* the existence of THE.[16] Alternatively, the so-called THE responses may arise from the overlapping of two distinct AHE contributions, such as $SrRuO_3$ systems and TI systems.[16-21] Note that in these previously reported TI systems, the sign of AHE did not remain the same with varying gate voltage and temperature;[19-21] nevertheless, the AHE showed no sign change with gate voltage and temperature in our top-gated BST on EuIG heterostructure. Thus, we have conducted the Hall measurements using the minor loop method to clarify the origins of the THE responses.[21] For the artificial THE-like responses, which result from two overlapping AHE loops, the Hall traces with different sweeping magnetic fields will not follow the trajectory of the full loop.[21] For the genuine THE signals, they will coincide within the full loop, which is the case of the results shown in Figure S3.

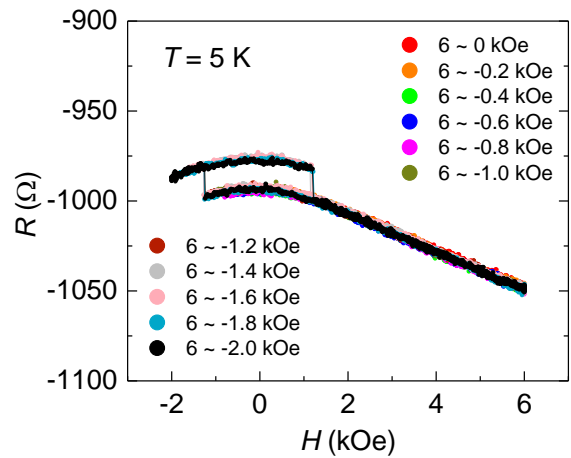

**Figure S3.** Minor loops at 5 K vs magnetic field with systematically increased sweeping range.

**Section S3. Discussion on the possibility of Hall effect data from two distinct AHE contributions**

Given the debates discussed in Section S2 in Supporting Information, we further demonstrate our experimental results and fitting curves in Figures S4a, b, where the fitting curves are composed of two AHE loops, $R_{AHE} = R_{AHE1-max}\ tanh\left(\frac{H \pm H_{c1}}{H_{01}}\right) + R_{AHE2-max}\ tanh(\frac{H \pm H_{c2}}{H_{02}})$. By applying a more positive gate voltage, the hysteretic contribution with a positive sign (AHE2) disappeared at $V_g \sim V_{CNP}$, and then the sign of AHE2 changed to negative at $V_g > V_{CNP}$. As summarized in Figure S4e, AHE1 did not change the sign with the gate voltage; however, AHE2 changed the sign with the gate voltage. In addition, we further analyzed the temperature-dependent results. As shown in Figure S4f, the AHE2 gradually diminished with increasing temperatures and disappeared at 75 K; in contrast, the AHE1 still existed up to 300 K (not shown).

The coercive field of the AHE1 remained nearly the same with varying gate voltage, which was expected and was also demonstrated in our previous work.[2] The coercive field only depends on the thermal activation of the domain walls in EuIG, thus expected to be independent of the gate bias, namely the $E_F$ of BST. However, the coercive field of AHE2 varied with gate voltages and did not decrease with increasing temperatures. (see Figures S4g, h) Since the gate- and temperature-dependent coercive field of AHE2 could not be explained physically, the excessive Hall signals on the AHE loops in this work were less likely to result from the two distinct AHE contributions.

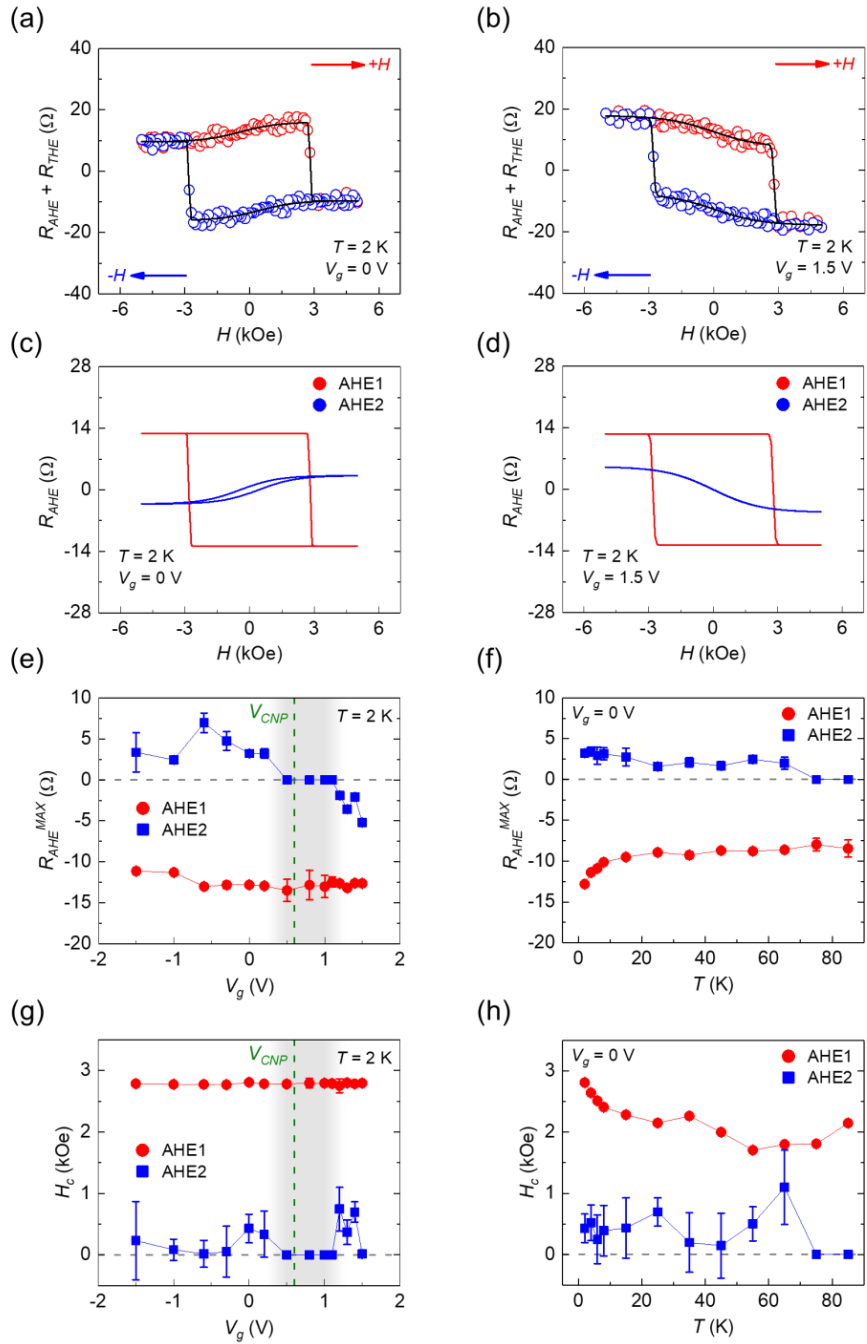

**Figure S4.** (a, b) The scattered points are the Hall data after subtracting the linear OHE background; the solid lines are the fitted curves. The fitting curves in (a) and (b) are composed of (c) two AHE

loops with the opposite signs and (d) two AHE loops with the same sign, respectively. The gate and temperature dependences of the $R_{AHE}$ in AHE1 and AHE2 are shown in (e) and (f), respectively. The gate and temperature dependences of the $H_c$ in AHE1 and AHE2 are shown in (g) and (h), respectively.

**Section S4. Discussion on the rationale of subtracting a square hysteresis for the AHE loop to derive THE**

From the discussion in Sections S2 and S3 in Supporting Information, we now confirm that the Hall signals were the sum of the ordinary Hall component, one hysteresis loop from AHE, and an additional contribution from THE. In general, the component from the OHE is a linear background. However, the shape and absolute values of the AHE contribution could affect the shape and values of THE signals in the data processing.

A viable way of deriving the AHE contribution is by measuring the magnetic hysteresis loop of the sample. The AHE contribution should be proportional to the magnetization hysteresis. Nevertheless, in the BST/EuIG/GGG, the paramagnetic contribution of the GGG substrate became extremely large at low temperatures, making the magnetization measurement very difficult in using traditional magnetometers, such as superconducting quantum interference device (SQUID) magnetometer and vibrating-sample magnetometer (VSM). Therefore, in the current work, the AHE contribution was fitted by using a hyperbolic tangent function to mimic the square hysteresis of AHE.

In the following, we elaborate on the rationale that the MPE-induced AHE loop is likely of a square, not other shapes. In the previous work done by M.-X. Guo *et al*., the *out-of-plane* magnetic hysteresis loops of EuIG were square at room temperature.[1] Square AHE loops were observed in multiple samples with BST on top of EuIG, as reported by W.-J. Zou *et al.*[2] Also, a square magnetic hysteresis was confirmed by AGM in 4 nm and 7 nm thick BST/EuIG at room temperature. (see Figure S2 in Supporting Information) As the temperature decreased, square AHE loops were retained from 400 K to 10 K in 7 nm thick BST/EuIG, in which no THE was observed.[2] Non-square like loops were only observed in 4 nm thick BST/EuIG below 75 K; the loops were all square above 75 K. Thus, it is reasonable to expect that the AHE loops at all temperatures were square, and the non-squared loops were the superposition of AHE and THE at temperatures below the critical temperature for THE, which is 75 K in this work.

Finally, we discuss one hypothetical case that the shape of AHE contribution is not a square hysteresis at low temperatures, but of some unusual shape. As long as the sign of the AHE loop remains independent of gate bias, subtracting the AHE contribution from the Hall measurement data is unlikely to result in an extra sign-change feature to the data presented in Figure 3. Therefore, we expect that the essential sign-reversal feature of THE will still be retained, similar to the data shown in Figure 3c. Given the considerations above, we have adopted the subtraction of square AHE hysteresis loops from the $R_{AHE} + R_{THE}$ data to deduce the THE signals for all temperatures.

## Section S5. Comparison of THE in various systems hosting magnetic skyrmions

**Table S1.** THE magnitude, temperature window for THE, and maximal in-plane magnetic field before the disappearance of THE in various systems hosting magnetic skyrmions.

|  | Maximal THE resistance ($\Omega$)[a] | Maximal THE resistivity ($\mu\Omega \cdot cm$) | Temperature window for THE (K) | Maximal in-plane field (kOe)[b] |
|---|---|---|---|---|
| TI-based system ||||  |
| BST/EuIG/GGG(001)* | 10 (2 K) | 4 (2 K) | 2–65 | 2.4 (5 K) |
| Cr-doped BST/BST/InP(111)[22] | 140 (2 K) |  | 2–18 |  |
| BST/MnTe/CrSe/GaAs(111)B[23] | 1.4 (1.9 K) |  | 1.9–20 |  |
| $Cr_2Te_3/Bi_2Te_3/Al_2O_3(0001)$[24] |  | 0.37 (18 K) | 2–20 |  |
| $Bi_2Se_3/BaFe_{12}O_{19}/Al_2O_3(0001)$[14] |  | 0.23 (2 K) | 2–80 | > 3.4 (2 K) |
| $CrTe_2/Bi_2Te_3/Al_2O_3(0001)$[15] |  | 1.39 (10 K) | 10–100 |  |
| Non-TI system ||||  |
| $Mn_{0.96}Fe_{0.04}Si/Si(111)$[25] |  | 0.022 (20 K) | 10–25 | 7.6 (20 K) |
| $EuO/YAlO_3(110)$[26] |  | 12.7 (50 K) | 2–50 | 2.3 (30 K) |
| $Pt/TmIG/Nd_3Ga_5O_{12}$[27] |  | 0.005 (370 K) | 360–410 | 0.05 (360 K) |
| $SrRuO_3/SrTiO_3(001)$[28] |  | 0.18 (10 K) | 10–60 | 64.9 (10 K) |

*The values presented in this work

[a]For comparison with TI-based systems hosting the insulating bulk state

[b]The maximal in-plane field before the disappearance of THE

## Section S6. Temperature-dependent THE with $V_g$ of 0 V

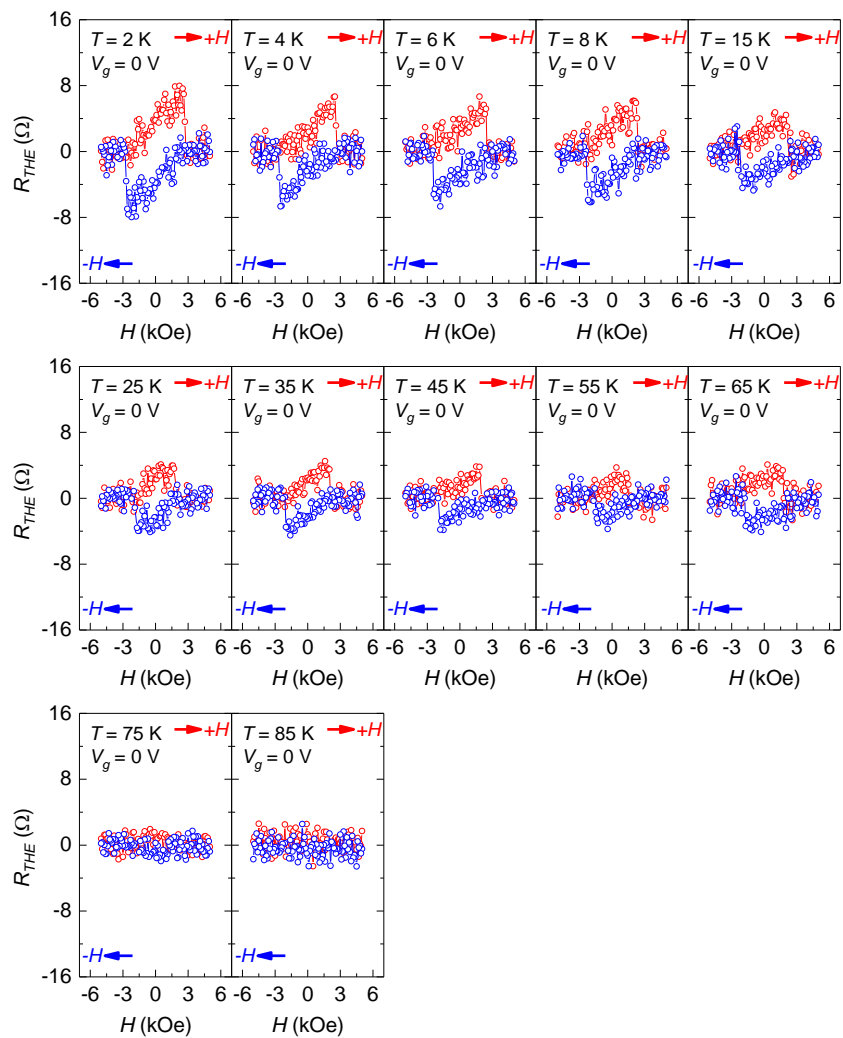

**Figure S5.** Temperature dependence of THE from 2 K to 85 K with $V_g$ of 0 V.

## Section S7. Gate-dependent THE at 2 K

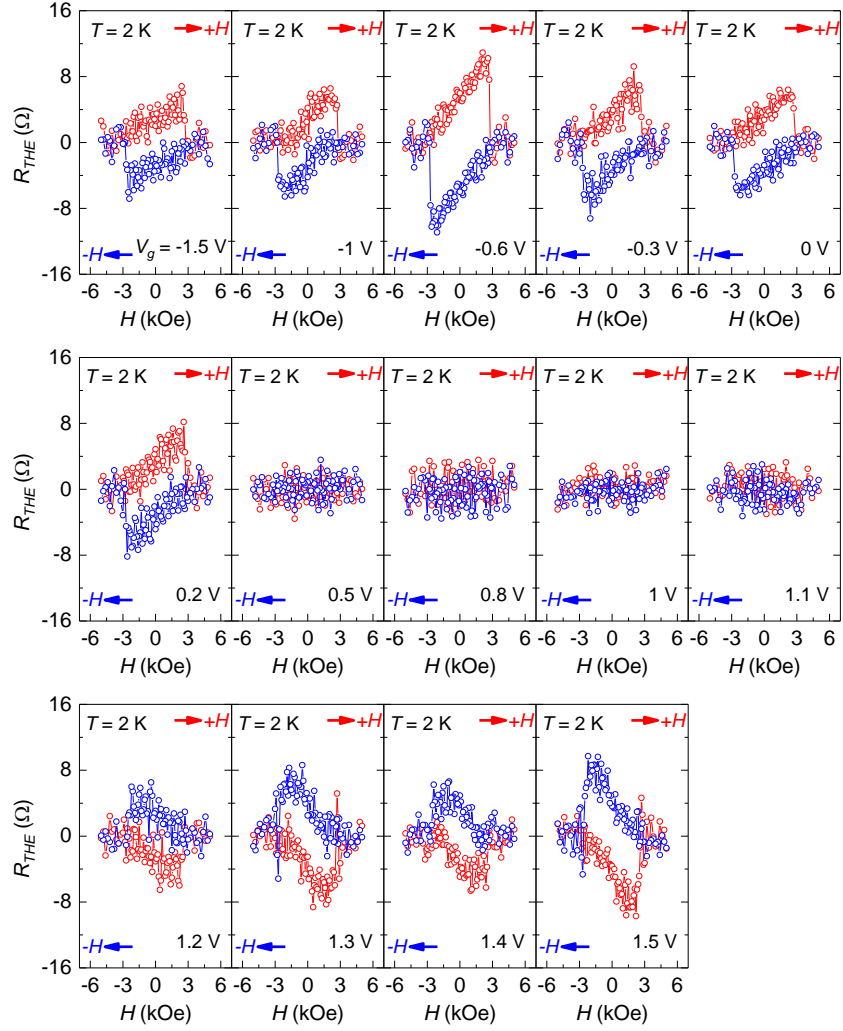

**Figure S6.** Gate dependence of the THE at 2 K with systematically varying $V_g$.

**Section S8. Estimate of $E_F$ manipulated by $V_g$**

By considering the conduction mainly from the TSS of the bulk-insulating BST, we calculated the modulation of the sheet carrier density ($n_{2D}$) per volt to be $\frac{4.30 \times 10^{12} \text{ cm}^{-2} \ (n_{2D} \text{ at } V_g = -1.5 \text{ V})}{|-1.5 \text{ V} - 0.6 \text{ V} \ (V_{CNP})|} \approx 2.05 \times 10^{12} \text{ cm}^{-2}\text{V}^{-1}$, where $n_{2D}$ is derived from the OHE background using $n_{2D} = \frac{1}{R_H e}$. Since the magnetic gap is small, we here assume the band dispersion of the gapped Dirac cone is very similar to that of the gapless Dirac cone with nearly the same slope. In a 2D system, the Fermi wavevector ($k_F$) can be derived from the equation $k_F = \sqrt{4\pi n_{2D}}$. By knowing the $k_F$, the $E_F$ can be further extracted from the linear dispersion of the Dirac cone with the equation $E_F = \hbar v_D k_F$, where the Dirac velocity $v_D \sim 3.76 \times 10^5$ m/s is referred to the data of $(Bi_{0.25}Sb_{0.75})_2Te_3$ in the literature.[29] Hence, we roughly estimated that the $E_F$ tuning *via* the gate bias in Figure 3 was from 128 meV ($V_g$ = -1.5 V) below the gapped Dirac point to 84 meV ($V_g$ = +1.5 V) above the gapped Dirac point. The tunability of $E_F$ might be overestimated because the $n_{2D}$ at $V_g$ = -1.5 V is at the margin of the ambipolar transport region.

## Section S9. Gate-dependent results of one additional field-effect device

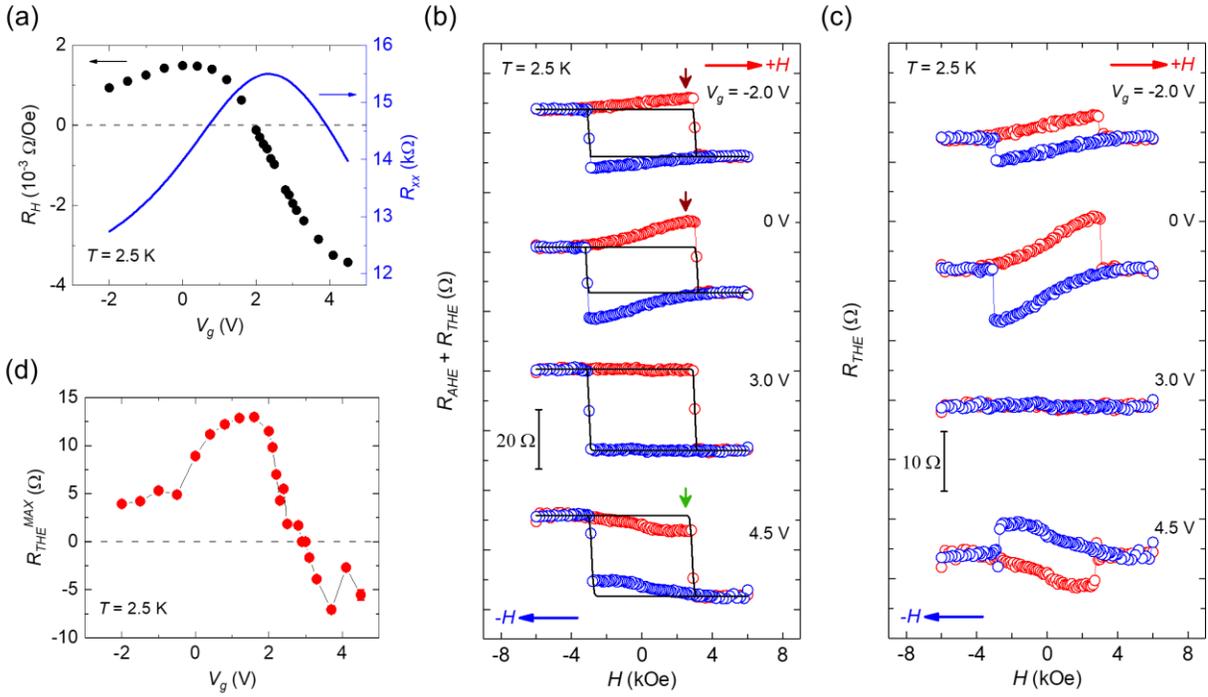

**Figure S7.** Additional data from another top-gated 4 nm $(Bi,Sb)_2Te_3$/20 nm EuIG at 2.5 K. Gate dependences of (a) $R_H$ (left) and $R_{xx}$ (right), (b) $R_{AHE} + R_{THE}$ data, (c) $R_{THE}$, and (d) $R_{THE}^{MAX}$.

**Section S10. Clarification to highlight the difference between this work and the previous gate-dependent THE results of W.-J. Zou *et al.***

In the previous work of our group by W.-J. Zou *et al.*, THE-like features were found and coexisted with AHE at low temperatures as the BST thickness was decreased to 4 nm.[2] By applying $V_g$ larger than $V_{CNP}$ (-1.4 V), the shape changed in the hysteresis curves were found in the preliminary Hall data without detailed analysis. (see scattered points in Figure S8)

After conducting more experiments as presented in this work, we confirmed that these shape changes did not occur by accident, but had their physical origin. To clarify the difference between the previous results by W.-J. Zou *et al.* and this work,[2] we re-analyzed the gate-dependent THE and $R_{THE}^{MAX}$ of the former, as shown in Figures S9 and S10, respectively. The consistency of these two sets of results is found. The sign of THE changed when the charge carriers were altered from electrons to holes. Notice that the feature of two $H_c$ in the former work was not observed in the present work. The reason for that is so far unclear.

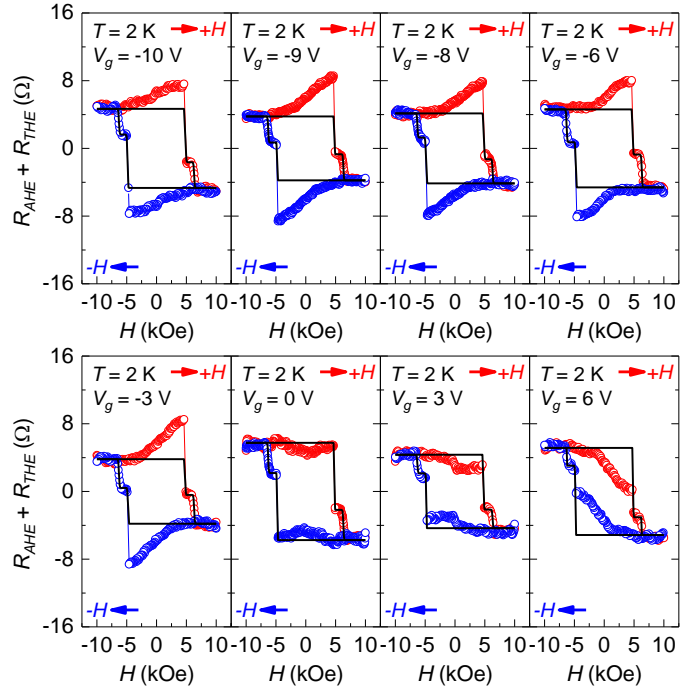

**Figure S8.** Gate-dependent $R_{AHE} + R_{THE}$ data of the previous work by W.-J. Zou *et al.*[2]

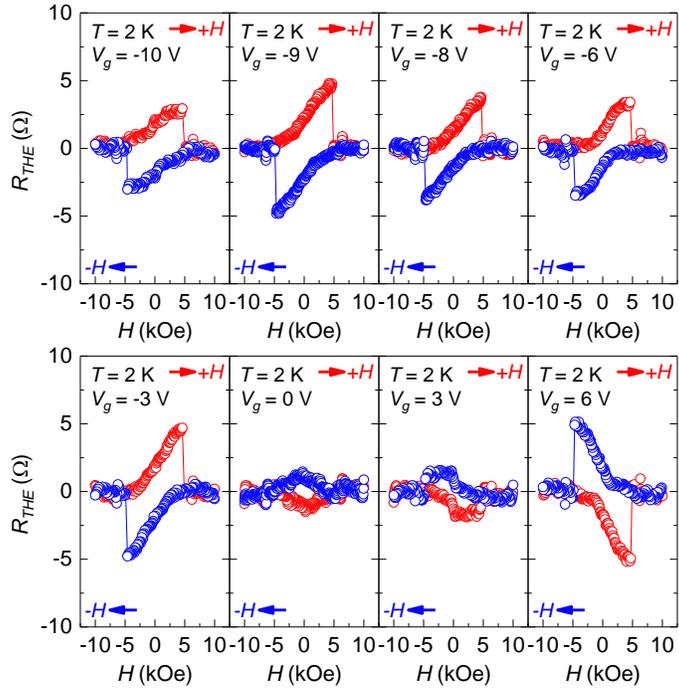

**Figure S9.** Gate-dependent $R_{THE}$ of the previous work by W.-J. Zou *et al.*[2]

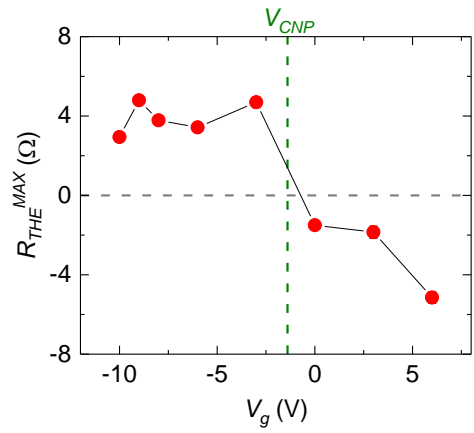

**Figure S10.** Gate dependence of $R_{THE}^{MAX}$ of the previous work by W.-J. Zou *et al.*[2]